\documentclass[article,shortnames,nojss]{jss}

\usepackage{natbib}
\usepackage{amsmath}
\usepackage{rotating}
\usepackage{url}

\newcommand\T{\rule{0pt}{2.6ex}}

\def\mbf#1{{%         \mbf{X} makes X a math bold letter
\mathchoice%          selects with respect to current style
{\hbox{\boldmath$\displaystyle{#1}$}}%      case 1 is displaystyle
{\hbox{\boldmath$\textstyle{#1}$}}%         case 2 is textstyle
{\hbox{\boldmath$\scriptstyle{#1}$}}%       case 3 is scriptstyle
{\hbox{\boldmath$\scriptscriptstyle{#1}$}}% case 4 is scriptscriptstyle
}}
\def\vec{\mbf}

\author{Julian Taylor\\ University of Adelaide \And David Butler \\ Queensland Government}
\title{\proglang{R} Package \pkg{ASMap}: Efficient Genetic Linkage Map
  Construction and Diagnosis}

\Plainauthor{Julian Taylor, David Butler} %% comma-separated
\Plaintitle{R Package ASMap: Efficient Genetic Linkage Map
  Construction and Diagnosis} %% without formatting
\Shorttitle{Linkage Map Construction and Diagnosis} %% a short title (if necessary)

\Abstract{
Although various forms of linkage map construction software are widely
available, there is a distinct lack of packages for use in
the \proglang{R} statistical computing environment \citep{rsoft15}. This article
introduces the \pkg{ASMap} linkage map construction \proglang{R} package which
contains functions that use the efficient MSTmap algorithm \citep{mst08} for clustering and
optimally ordering large sets of markers. Additional to the
construction functions, the package also contains a suite of tools
to assist in the rapid diagnosis and repair of a constructed linkage
map. The package functions can also be used for post
linkage map construction techniques such as fine mapping or combining maps of the
same population. To showcase the efficiency and functionality of \pkg{ASMap},
the complete linkage map construction process is demonstrated with a
high density barley backcross marker data set.
}

\Plainkeywords{linkage map construction, genetics, quantitative trait loci, R} %% without formatting
\Address{
  Julian Taylor\\
  School of Agriculture, Food \& Wine\\
  The University of Adelaide\\
  PMB 1, Glen Osmond, SA, 5064, Australia\\
  E-mail: \email{julian.taylor@adelaide.edu.au}\\
\\
  David Butler\\
  Department of Agriculture, Fisheries \& Forestry\\
  Queensland Government\\
  PMB 102, Toowoomba, QLD, 4350, Australia\\
  E-mail: \email{david.butler@daf.qld.gov.au}\\
}

\setkeys{Gin}{width=.97\textwidth}

\begin{document}

\section{Introduction}

Genetic linkage maps are widely used in the biological research
community to explore the underlying DNA of populations. They generally
consist of a set of polymorphic genetic markers spanning the entire
genome of a population generated from a specific
cross of parental lines. This exploration may involve the dissection
of the linkage map itself to understand the genetic landscape of the
population or, more commonly, it is used to conduct gene-trait
associations such as quantitative trait loci (QTL) analysis or genomic
selection (GS). For QTL analysis, the interpretation of significant
genomic locations is enhanced if the linkage map contains markers that
have been assigned and optimally ordered within chromosomal
groups. This can be achieved algorithmically by using linkage map
construction techniques that utilise the laws of Mendelian genetics
\citep{stur13}.

In the relatively short history of linkage map construction there has
been an abundance of stand alone software. Early linkage map
construction algorithms used brute force
combinatoric methods to determine marker order \citep{lg87, lw89} and were built into
historical versions of popular software packages \pkg{Mapmaker}
\citep{land87} and \pkg{JoinMap} \citep{stam93}. As the number of markers increased
non-combinatoric methods emerged, including SERIATION \citep{ser87} and
Rapid Chain Delineation (RCD) \citep{doe96}. A version of RCD was
implemented in the intuitive graphically oriented package \pkg{Map Manager}.
\citep{man01}. Although \pkg{Map Manager} enjoyed a meteoric rise in
use, the implemented RCD algorithm was sub-optimal and the software was limited
to reduced numbers of markers. \cite{car97} and \cite{liu98}
recognised a more efficient approach
to marker ordering could be attained from finding solutions to the
travelling salesman problem. Computational methods that exploited this
knowledge were quickly adopted in construction algorithms including the
Evolution-Strategy algorithm \citep{mest03, mest03a} implemented in
\pkg{Multipoint} and the RECORD algorithm \citep{rec05}
available as command line software and later implemented in
software packages \pkg{IciMapping} \citep{meng15} and
\pkg{onemap} \citep{one15}. Other construction algorithm variants
involving efficient solutions of the travelling salesman problem
included the unidirectional growth (UG)
algorithm \citep{tf06}, AntMap algorithm \citep{in06}, the
MSTmap algorithm \citep{mst08} and Lep-MAP
\citep{ras13}. Unfortunately these more recent algorithms have only
been available as downloadable low-level source code that require
compilation before use.

In the \proglang{R} statistical computing environment
there were only two packages available for linkage map
construction. The \pkg{qtl} package \citep{br15, brosen09} has matured
considerably since it inception in 2001 and provides users with a suite of functions
for linkage map construction and QTL analysis across a wide set of
populations. Unfortunately, the marker ordering algorithms in the
package rely on computationally cumbersome combinatoric methods and
multiple stages to achieve optimality. In a more recent addition to
the \proglang{R} contributed package list, \pkg{onemap}  \citep{one15}
contains a collection of linkage map construction tools specialized for a
restricted set of inbred and outbred populations. The package implements well known marker
ordering algorithms including SERIATION \citep{ser87}, RECORD
\citep{rec05}, RCD \citep{doe96} and UG \citep{tf06}. Unfortunately
these algorithms have been implemented in native \proglang{R} code and
lack the computational expediency of the low-level source code equivalents.

In an attempt to circumvent these computational issues we developed
the \pkg{ASMap} package \citep{tb15} which is freely downloadable at
\url{http://CRAN.R-project.org/package=ASMap}. The package contains linkage map
construction functions that fully utilize the freely available \proglang{C++}
source code (\url{http://alumni.cs.ucr.edu/~yonghui/mstmap.html}) for
the MSTmap algorithm derived in \cite{mst08} and outlined in
Section \ref{sec:mst}. The algorithm uses the minimum spanning tree of
a graph \citep{ct76} to cluster markers into
linkage groups as well as find the optimal marker order within each linkage group.
In contrast to \pkg{qtl} and \pkg{onemap}, genetic linkage maps are
constructed very efficiently without the requirement for multiple
ordering stages. The algorithm is restricted to linkage map
construction with Backcross (BC), Doubled Haploid (DH) and Recombinant Inbred (RIL)
populations. For RIL populations, the level of self pollination can
also be given and consequently the algorithm can handle selfed F2, F3,
$\ldots$, F$r$ populations where $r$ is the level of selfing. Advanced
RIL populations $(r \rightarrow \infty)$ are also allowed and are treated similar to a DH
population for the purpose of linkage map construction.

An overview of the \pkg{ASMap} package functions are presented in
Section \ref{sec:asmap}. Two linkage map construction functions are available
that provide users with a fast, flexible set of tools for
constructing linkage maps using the MSTmap algorithm. To
complement the efficient construction functions the package also
contains a function that pulls markers of different types from the
linkage map, temporarily placing them
aside. This is complemented by a function that pushes markers back to
linkage groups at any stage of the construction or
reconstruction process. To efficiently diagnose the quality of the constructed
map the package contains graphical functions that simultaneously display multiple panel
profiles of linkage map statistics associated with the genotypes or
marker/intervals. Where possible, the \pkg{ASMap} package
uses the \code{"cross"} object format (see the \pkg{qtl} function
\code{read.cross()} for the structure of its genetic objects. Once the
class of the object is appropriately set, both \pkg{ASMap} and
\pkg{qtl} functions can be used synergistically to construct, explore
and manipulate the object. To showcase the functionality of the
\pkg{ASMap} package, Section \ref{sec:work} presents an illustrative example
that involves the complete linkage map construction process for a
barley Backcross population. This section concludes with a short
summary on the use of \pkg{ASMap} functions in post construction linkage map development
techniques such as fine mapping and combining maps. The performance of
the MSTmap algorithm in \pkg{ASMap} is outlined in Section
\ref{sec:lim} and the article concludes with a short summary.

\section{MSTmap algorithm}
\label{sec:mst}

Following the notation of \cite{mst08} consider a Doubled
Haploid population of $n$ individuals genotyped across a set of $t$
markers where each $(j,k)$th entry of the $n \times t$ matrix $\vec{M}$
is either an $A$ or a $B$ representing the two parental
homozygotes in the population. Let $\vec{P}_{jk}$ be the probability
of a recombination event between the markers $(\vec{m}_j,
\vec{m}_k)$ where $0 \leq \vec{P}_{jk} \leq 0.5$. MSTmap uses two
possible weight objective functions based on recombination
probabilities between the markers
\begin{align}
  w_p(j,k) &= \vec{P}_{jk} \label{eq:weight1}\\
  w_{ml}(j, k) &= -\bigl(\vec{P}_{jk}\log\,\vec{P}_{jk} + (1 -
  \vec{P}_{jk})\log(\,1 - \vec{P}_{jk})\bigr) \label{eq:weight2}
\end{align}
In general $\vec{P}_{jk}$ is not known and so it is replaced by an
estimate, $d_{jk}/n$ where $d_{jk}$ corresponds to the hamming
distance between $\vec{m}_j$ and $\vec{m}_k$ (the number of
non-matching alleles between the two markers). This estimate, $d_{jk}/n$,
is also the maximum likelihood estimate for $\vec{P}_{jk}$ for the two
weight functions defined above.

\subsection{Clustering}
\label{sec:clust}

\begin{figure}
  \centering
\includegraphics[width = 11cm]{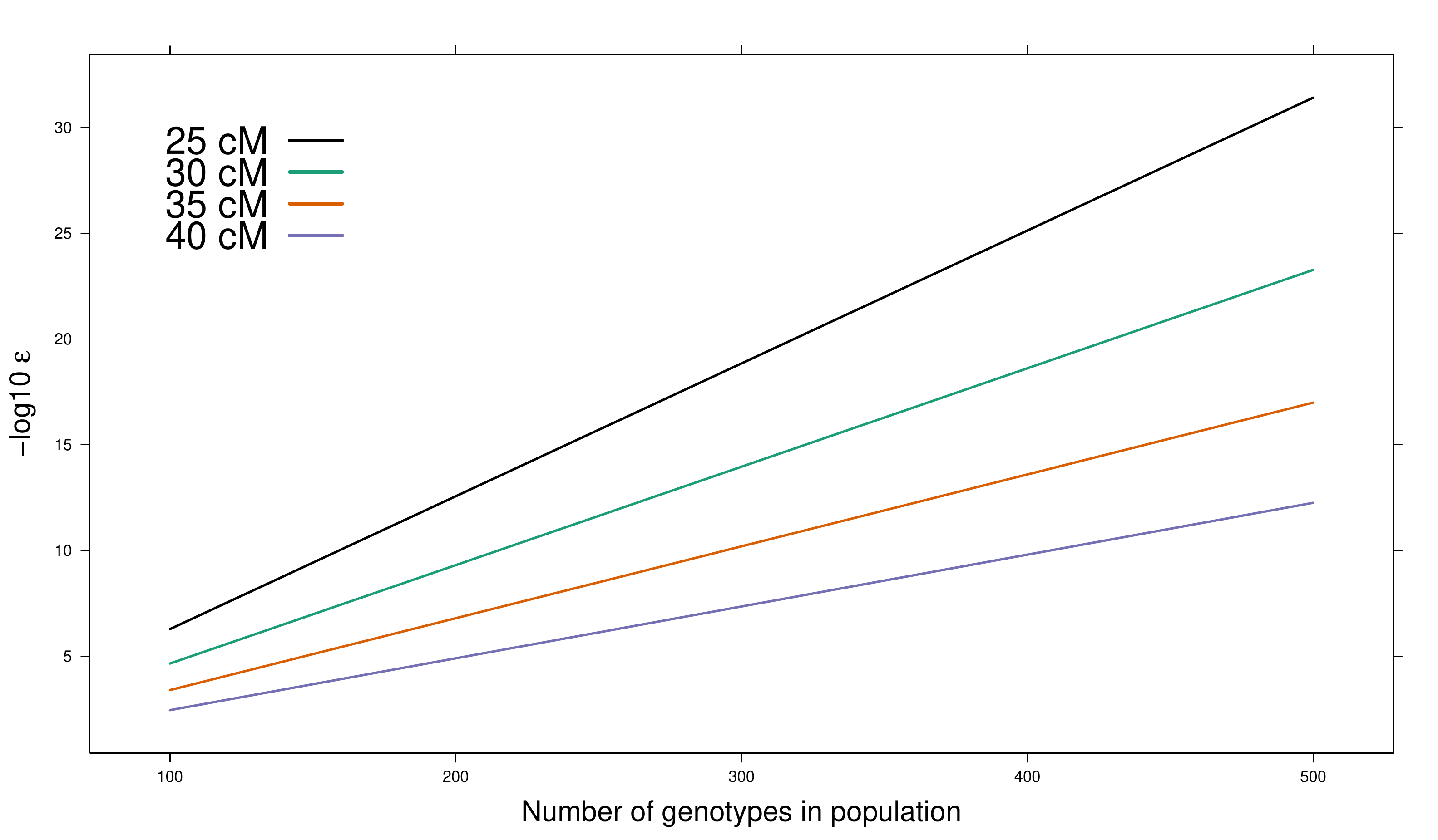}
\caption{Negative log10 $\epsilon$ versus the number of genotypes in
  the population for four threshold cM distances.}
\label{fig:threshold}
\end{figure}

If markers $\vec{m}_j$ and $\vec{m}_k$ belong to two different linkage
groups then $\vec{P}_{jk} = 0.5$ and the hamming distance between them
has the property $E(d_{jk}) = n/2$. Using these definitions, MSTmap determines whether markers
belong to the same linkage group using Hoeffdings inequality
\begin{align}
 P(d_{jk} < \delta) \leq \mbox{exp}(-2(n/2 - \delta)^2/n)
 \label{eq:hoeff}
 \end{align}
for $\delta < n/2$. For a given $P(d_{jk} < \delta) = \epsilon$ and
$n$, the equation $-2(n/2 - \delta)^2/n = \mbox{log}\,\epsilon$ is
solved to determine an appropriate hamming distance threshold,
$\hat{\delta}$. \cite{mst08} indicate that the choice of $\epsilon$ is not
crucial when attempting to form linkage groups. However, the equation
that requires solving is highly dependent on the number of individuals
in the population. For example, for a DH population, Figure
\ref{fig:threshold} shows the profiles of the negative $\log 10 \epsilon$ against
the number of individuals in the population for four threshold minimum
cM distances $(25, 30, 35, 40)$. MSTmap uses a default of $\epsilon = 0.00001$ which would work
universally well for population sizes of $n \sim 150\, \mbox{to}\, 200$. For larger
numbers of individuals, for example $n = 350$, the plot indicates an
$\epsilon = 10^{-12}$ to $10^{-15}$ would use a conservative minimum threshold of 30-35
cM before linking markers between clusters. If the default $\epsilon =
10^{-6}$ is given in this instance this threshold is dropped to $\sim 45$
cM and consequently distinct clusters of markers will appear
linked. For this reason, Figure \ref{fig:threshold} should always initially
be checked before linkage map construction to ensure an appropriate p-value is
given to the MSTmap algorithm.

To cluster the markers MSTmap uses an edge-weighted
complete graph for $\vec{M}$ where the individual markers are
vertices and the edges between any two markers $\vec{m}_j$ and
$\vec{m}_k$ is the pairwise hamming distance $d_{jk}$. Edges
with weights greater than $\hat{\delta}$ are then removed. The remaining
connected components allow the marker set $\vec{M}$
to be partitioned into $r$ linkage groups, $\vec{M} =
[\vec{M}_1,\ldots,\vec{M}_r]$.

\subsection{Marker ordering}
\label{sec:ord}

For simplicity, consider the $n \times t$ matrix of markers $\vec{M}$
belongs to the same linkage group. Preceding marker ordering, the
markers are ``binned'' into groups where, within each group,
the pairwise distance between any two markers is zero. The markers within
each group have no recombinations between them and are said to be
co-locating at the same genomic location for the $n$ genotypes used to
construct the linkage map. A representative marker is then chosen from each
of the bins and used to form the reduced $n \times t^*$ marker set
$\vec{M}^*$.

For the reduced matrix $\vec{M}^*$, consider the complete set of entries $(j, k)
\in (1, \ldots, t^*)$ for either weight function (\ref{eq:weight1}) or
(\ref{eq:weight2}). These complete set of entries can be viewed as the
upper triangle of a symmetric weight matrix $\vec{W}$. MSTmap views all
these entries as being connected edges in an undirected graph where
the individual markers are vertices. A marker order for the set $\vec{M}^*$,
also known as a travelling salesman path (TSP), is found by visiting
each marker once and summing the weights from
the connected edges. To find a minimum weight (TSP$_{min}$), MSTmap uses a minimum
spanning tree (MST) algorithm \citep{ct76}, such as Prims algorithm
\citep{prim57}. If the TSP$_{min}$ is unique then the MST is the
optimal order for the markers. For cases where the data contains
genotyping errors or lower numbers of individuals the MST may not be a
complete path and contain markers or small sets of markers as
individual nodes connected to the path. In these cases, MSTmap uses the
longest path in the MST as the backbone and employs several efficient local
optimization techniques such as K-opt, node-relocation and
block-optimize \citep[see][]{mst08} to improve the current minimum
TSP. By integrating these local optimization techniques into the
algorithm, MSTmap provides users with a true one stage marker ordering
algorithm.

A unique feature of the MSTmap algorithm is the use of an
Expectation-Maximisation (EM) type algorithm for the imputation of
missing allele scores that is tightly integrated with the ordering
algorithm for the markers. To achieve this the marker matrix $\vec{M}^*$ is converted to a
matrix, $\vec{A}$, where the entries represent the
probabilistic certainty of the allele being AA. For the $j$th marker and
$i$th individual then
\begin{align}
  \vec{A}(i,j) = \left\{\begin{array}{ll}
          1 & \quad \mbox{if $\vec{M}^*(i,j)$ is the A allele}\\
          0 & \quad \mbox{if $\vec{M}^*(i,j)$ is the B allele}\\
          \vec{R}_j/(\vec{R}_j + \vec{S}_j) & \quad \mbox{if
            $\vec{M}^*(i,j)$ is missing} \end{array}\right.\label{eq:rf}
\end{align}
where $\vec{R}_j = (1 - \hat{\vec{P}}_{j - 1,j})(1 -
\hat{\vec{P}}_{j,j+1})$, $\vec{S}_j = \hat{\vec{P}}_{j-1,j} \hat{\vec{P}}_{j,j+1}$
and $\hat{\vec{P}}_{j,j-1}, \hat{\vec{P}}_{j,j+1}$ are
estimated recombination fractions between the $(j-1)$th and $j$th
marker and $j$th and $(j+1)$th marker respectively. If
$\vec{M}^*(i,j)$ is missing then the equation on
the right hand side is the posterior probability of the missing value in
marker $j$ being the A allele for genotype $i$ given the current
estimate. Unlike the flanking marker methods of \citet{mc92,
  mc94} this equation represents a probabilistic approach to
imputation. The ordering algorithm begins by initially calculating
pairwise normalized distances between all markers in
$\vec{M}^*$ and deriving an initial weight matrix, $\vec{W}$. An
undirected graph is formed using the markers as vertices and the upper triangular
entries of $\vec{W}$ as weights for the connected edges. An MST of the undirected
graph is then found to establish an initial order for the
markers of the linkage group. For the current order at the $(j - 1, j,
j + 1)$th markers the E-step of algorithm requires updating the
missing observation at marker $j$ by updating the estimates
$\hat{\vec{P}}_{j-1,j} = \hat{d}_{j-1,j}/n$ and $\hat{\vec{P}}_{j,j+1} =
    \hat{d}_{j,j+1}/n$ in (\ref{eq:rf}). The M-step then re-estimates the
    pairwise distances between all markers in $\vec{M}^*$ where, for
    the $j$th and $k$th marker, is
\begin{align*}
\hat{d}_{jk} = \sum_{i = 1}^{t^*}\vec{A}(i,j)(1 - \vec{A}(i, k)) + \vec{A}(i,k)(1 - \vec{A}(i, j))
%\label{eq:dist}
\end{align*}
and the weight matrix $\vec{W}$ is recalculated. An undirected graph
is formed with the markers as vertices and the upper triangular
entries of $\vec{W}$ as weights for the connected edges. A new order of the markers is
derived by obtaining an MST of the undirected graph and the algorithm is repeated to
convergence. Although this requires several iterations to converge,
the computational time for the ordering algorithm remains
efficient.

If required, the MSTmap algorithm also detects and removes genotyping
errors as well as integrates this process into the ordering
algorithm. The technique involves using a weighted average of nearby
markers to determine the expected state of the allele. For individual
$i$ and marker $j$ the expected value of the allele is calculated
using
\begin{align*}
\mbox{E}[\vec{A}(i, j)] = \sum_{j \neq
    k}d_{j,k}^{-2}\vec{A}(i,k)\bigg/\sum_{j \neq k}d_{j,k}^{-2}
%\label{eq:err}
\end{align*}
In this equation the weights are the inverse square of the distance from
marker $j$ to its nearby markers. MSTmap only uses a small set of
nearby markers during each iteration and the observed allele is
considered suspicious if $|\mbox{E}[\vec{A}(i, j)] - \vec{A}(i, j)| >
0.75$.  If an observation is detected as suspicious it is treated as
missing and imputed using the EM algorithm discussed previously. The
removal of the suspicious allele has the effect of reducing the number
of recombinations between the marker containing the suspicious observation and the
neighbouring markers. This has an influential effect on
the genetic distance between markers and the overall length of the
linkage group.

The complete algorithm used to initially cluster the markers into
linkage groups and optimally order markers within each linkage group,
including imputing missing alleles and error detection, is known as
the MSTmap algorithm.

\subsection{Extension to RIL populations}

The MSTmap algorithm can also be used to construct linkage maps for
RIL populations generated through self-pollination of F1 derived
individuals. These include inbred F2, F3, $\ldots$, F$r$
populations, where $r$ is the level or generation of selfing as well
as Advanced RIL populations created by $r \rightarrow \infty$
levels of selfing. Non-advanced RIL populations contain three distinct genotypic
states, two parental homozygotes (AA, BB) in equal proportions and a
heterozygote (AB) with expected proportion determined by
the simple decaying equation $2^{-(r - 1)}$. As $r \rightarrow \infty$
this expected fraction tends to zero and the population can be
considered to be an Advanced RIL containing parental (AA, BB) homozygotes
only.

To ensure the MSTmap algorithm can be efficiently used to perform
clustering and optimal ordering of markers for RIL populations,
accurate estimates of pairwise recombination probabilities between
markers are required. For an Advanced RIL the estimated recombination
probability between any two markers $\vec{m}_j$ and $\vec{m}_k$, can
be directly calculated using the result in \cite{bro05}, namely
\begin{align}
  \hat{\vec{P}}^*_{jk} = (\hat{\vec{P}}_{jk}/2)/(1 - \hat{\vec{P}}_{jk})
  \label{eq:trf}
\end{align}
where $\hat{\vec{P}}_{jk} = d_{jk}/n$ is the estimated recombination
probability between the two markers from a DH population. For any
non-Advanced RIL, $\vec{P}_{jk}^*$ cannot be directly calculated and
the MSTmap algorithm uses the recurrence
relation results of \cite{hw31} \citep[see Supplementary Text
S1][]{mst08} and a simple stepwise optimization procedure to
closely approximate $\vec{P}_{jk}^*$. Although Hoeffdings inequality
(\ref{eq:hoeff}) is well-defined for DH and BC populations, it is also used in the MSTmap
algorithm to cluster markers for RIL populations. For large enough
$r$, (\ref{eq:trf}) can be approximately used to investigate the threshold p-value
required for the inequality. At its theoretical boundaries, zero and 0.5, $\vec{P}_{jk}^*$
and $\vec{P}_{jk}$ are equivalent but for $\vec{P}_{jk}$ between
$0.25$ and $0.35$, $\vec{P}_{jk}^*$ is substantially decreased creating a
reduction of 10 cM+ in the genetic distance between the markers.
Figure \ref{fig:threshold} indicates that this 10 cM reduction would
require the p-value to be squared to achieve the appropriate
threshold for clustering markers in RIL populations. Optimal ordering of
markers within clustered linkage groups is then accomplished using the methods
outlined in Section \ref{sec:ord}. However, for non-Advanced RIL
populations only, the additional MSTmap algorithm features, including imputation
of missing alleles scores and the detection of potential genotyping errors, have
not been implemented. 

\newpage

\section[ASMap package]{\pkg{ASMap} package}
\label{sec:asmap}

\subsection{Map construction functions}
\label{sec:cons}

The \pkg{ASMap} package contains two linkage map construction functions
that allow users to fully utilize the MSTmap parameters listed at
\url{http://alumni.cs.ucr.edu/~yonghui/mstmap.html} and available for
use with the source code.

\begin{CodeInput}
mstmap.data.frame(object, pop.type = "DH", dist.fun = "kosambi",
     objective.fun = "COUNT", p.value = 1e-06, noMap.dist = 15,
     noMap.size = 0, miss.thresh = 1, mvest.bc = FALSE, detectBadData = FALSE,
     as.cross = TRUE, return.imputed = TRUE, trace = FALSE, ...)
\end{CodeInput}
The explicit form of the data frame \code{object} required for
\code{mstmap.data.frame()} is borne from the syntax of the marker file required for
using the MSTmap source code. It must have markers in rows and genotypes in
columns. Marker names are required to be in the \code{rownames} component of
the data frame, genotype names should reside in the \code{names} and each
of the columns of the data frame must be of class \code{"character"} (not
factors). The available populations that can be passed to the argument
\code{pop.type} are \code{"BC"} Backcross, \code{"DH"} Doubled
Haploid, \code{"ARIL"} Advanced Recombinant Inbred and
\code{"RILn"} Recombinant Inbred with $n$ levels of selfing. It is
recommended to set \code{as.cross = TRUE} to ensure the
returned object can be used with the suite of \pkg{ASMap} and \pkg{qtl} package functions.

The second function provides greater linkage map construction
flexibility by allowing users to pass an unconstructed or constructed
\code{"cross"} object created from package \code{qtl}.
\begin{CodeInput}
mstmap.cross(object, chr, id = "Genotype", bychr = TRUE, suffix = "numeric",
     anchor = FALSE, dist.fun = "kosambi", objective.fun = "COUNT",
     p.value = 1e-06, noMap.dist = 15, noMap.size = 0, miss.thresh = 1,
     mvest.bc = FALSE, detectBadData = FALSE, return.imputed = FALSE,
     trace = FALSE, ...)
\end{CodeInput}
The \code{object} needs to inherit from one of the allowable classes
available in the \pkg{qtl} package, namely
\code{"bc","dh","riself","bcsft"} where \code{"bc"} is a
Backcross \code{"dh"} is Doubled Haploid, \code{"riself"} is an advanced
Recombinant Inbred (see \code{?convert2riself}) and \code{"bcsft"}
is a Backcross/Self (see \code{?convert2bcsft}). The functions
flexibility stems from the appropriate use of \code{bychr} and \code{chr}
arguments. The logical flag \code{bychr = FALSE} ensures the
the subset of linkage groups defined by \code{chr} will be bulked
and reconstructed whereas \code{bychr = TRUE} confines the
reconstruction within each linkage group defined by \code{chr}.

Users need to be aware the \code{p.value} argument available for both
construction functions plays a crucial role in determining the
clustering of markers to distinct linkage groups. Section
\ref{sec:clust} shows the separation of marker groups is highly
dependent on the the number of individuals in the
population. As a consequence, some trial and error may be required to determine an
appropriate \code{p.value} for the linkage map being constructed. We
have also provided an additional feature to the construction functions
that allows the imputed probability matrix of representative markers
to be returned if \code{return.imputed = TRUE}.

\subsection{Pulling and pushing markers}
\label{sec:pp}

Often in linkage map construction some pruning of the markers occurs
before initial construction. For example, this may be the removal of markers with a
proportion of missing values higher than some desired threshold as
well as markers that are significantly distorted from their expected
Mendelian segregation patterns. The removal is usually permanent and the
possible importance of some of these markers may be overlooked. A
preferable system would be to identify and place the problematic markers aside with
the intention of checking their usefulness at a later stage of the
construction process. The \pkg{ASMap} package contains two functions that allow you to do
this.
\begin{CodeInput}
pullCross(object, chr, type = c("co.located","seg.distortion","missing"),
     pars = NULL, replace = FALSE, ...)
pushCross(object, chr, type = c("co.located","seg.distortion","missing",
     "unlinked"), unlinked.chr = NULL, pars = NULL, replace = FALSE, ...)
\end{CodeInput}
The construction helper functions share three types of markers that
can be ``pulled/pushed'' from linkage maps. These include markers that are co-located with
other markers, markers that have some defined segregation distortion
and markers with a defined proportion of missing values. If the
argument \code{type} is \code{"seg.distortion"} or \code{"missing"}
then the initialization function \code{pp.init()}
\begin{CodeInput}
pp.init(seg.thresh = 0.05, seg.ratio = NULL, miss.thresh = 0.1, max.rf =
     0.25, min.lod = 3)
\end{CodeInput}
is used to determine the appropriate threshold parameter setting
(\code{seg.thresh}, \code{seg.ratio}, \code{miss.thresh}) that will be
used to pull/push markers from the linkage map. Users can set their
own parameters by appropriate use of the \code{pars} argument. For each different \code{type},
\code{pullCross()} will pull markers from the map and
place them in separate elements of the returned object. Within the elements, vital
information is kept that can be accessed by \code{pushCross()} to push
the markers back at a later stage of linkage map construction. The
function \code{pushCross()} also contains another marker type
called \code{"unlinked"} which, in conjunction with the argument
\code{unlinked.chr}, allows users to push markers from an unlinked
linkage group in the \code{geno} element of the \code{object} into
established linkage groups. This mechanism becomes vital, for example,
when pushing new markers into an established linkage map.

\subsection{Visual diagnostics}
\label{sec:prof}

To provide a complete system for efficient linkage map construction
\pkg{ASMap} contains graphical functions for visuals diagnosis of your
constructed linkage map. Three flexible functions are provided and
examples of their use are given in Section \ref{sec:work}.

\begin{CodeInput}
profileGen(cross, chr, bychr = TRUE, stat.type = c("xo", "dxo", "miss"),
     id = "Genotype", xo.lambda = NULL, ...)
\end{CodeInput}
The function \code{profileGen()} calls \code{statGen()} to obtain statistics for graphically
profiling information about the genotypes across the marker set and also returns the
statistics invisibly after plotting. The current statistics that
can be calculated and profiled include
\begin{list}{$\bullet$}{\setlength{\itemsep}{1ex}\setlength{\parsep}{0ex}\setlength{\parskip}{0.5ex}}
\item \texttt{"xo"} : number of crossovers.
\item \texttt{"dxo"} : number of double crossovers.
\item \texttt{"miss"} : number of missing values.
\end{list}
The two statistics \code{"xo"} and \code{"dxo"} are only useful for
constructed linkage maps. From the authors experience, they represent
the most vital two statistics for determining a linkage maps
quality. Inflated crossover or double crossover rates of any
genotypes indicate problematic lines and should be
questioned. Significant crossover rates can be checked by manually
inputting a median crossover rate using the argument
\code{xo.lambda}. Additional graphical parameters can be passed to
the high level lattice function \code{xyplot()} through the
\code{"..."} argument
\begin{CodeInput}
profileMark(cross, chr, stat.type = "marker", use.dist = TRUE,
     map.function = "kosambi", crit.val = NULL, display.markers = FALSE,
     mark.line = FALSE, ...)
\end{CodeInput}
The function \code{profileMark()} calls \code{statMark()} and
graphically profiles marker/interval statistics as well as
returns them invisibly after plotting. The current
marker statistics that can be profiled are
\begin{list}{$\bullet$}{\setlength{\itemsep}{0.8ex}\setlength{\parsep}{0ex}\setlength{\parskip}{0.5ex}}
 \item \texttt{"seg.dist"}: -log10 p-value from a test of segregation distortion.
  \item \texttt{"miss"}: proportion of missing values.
  \item \texttt{"prop"}: allele proportions.
  \item \texttt{"dxo"}: number of double crossovers.
\end{list}
The current interval statistics that can be profiled are
\begin{list}{$\bullet$}{\setlength{\itemsep}{1ex}\setlength{\parsep}{0ex}}
  \item \texttt{"erf"}: estimated recombination fractions.
  \item \texttt{"lod"}: LOD score for the test of no linkage.
  \item \texttt{"dist"}: interval map distance.
  \item \texttt{"mrf"}: map recombination fraction.
  \item \texttt{"recomb"}: number of recombinations.
\end{list}
The function allows any combination of marker/interval statistics to be plotted
simultaneously on a multi-panel lattice display. There is a \code{chr}
argument to subset the linkage map to user defined linkage groups. If
\code{crit.val = "bonf"} then markers that have significant
segregation distortion greater than the family wide alpha level of
$0.05/m$, where $m$ is the number of markers, will be annotated in
marker panels. Similarly, intervals that have a significantly weak
linkage from a test of the recombination fraction of $r = 0.5$ will also be annotated in the
interval panels. All linkage groups are highlighted in a different colours to ensure
they can be identified clearly. The lattice panels ensure that marker
and interval statistics are seamlessly plotted together so problematic
regions or markers can be identified efficiently. Additional graphical
parameters can be passed to \code{xyplot()} through the \code{"..."}
argument.

\begin{CodeInput}
heatMap(x, chr, mark, what = c("both", "lod", "rf"), lmax = 12, rmin = 0,
     markDiagonal = FALSE, color = rev(colorRampPalette(brewer.pal(11,
        "Spectral"))(256)), ...)
\end{CodeInput}
\pkg{ASMap} contains an improved version of the heat map
that rectifies limitations of the heat map,
\code{plot.rf()}, available in \pkg{qtl}. The function independently
plots the LOD score on the bottom triangle of the
heat map as well as the actual estimated recombination fractions (RFs) on the
upper triangle. A colour key legend is also provided for the RFs and
LOD scores on the left and right hand side of the heat map
respectively. As the actual estimated RFs are plotted, the scale of
the legend includes values beyond the theoretical threshold of 0.5. By
increasing this scale beyond 0.5, potential regions where markers out
of phase with other markers can be recognised. Similar to
\code{plot.rf()}, the \code{heatMap()} function allows subsetting of
the linkage map by \code{chr} and users can further subset
the linkage groups using the argument \code{mark} by indexing a set of
markers within linkage groups defined by \code{chr}.

\subsection{Miscellaneous functions}
\label{sec:misc}

In the authors experience, the assumptions of how the individuals of a
population are genetically related is rarely checked throughout the
construction process. Too often unconstructed or constructed
linkage maps contain individuals that are closely related beyond the simple
assumptions of the population. \pkg{ASMap} contains a function for the detection and
reporting of the relatedness between individuals as well as a
function for forming consensus genotypes if genuine clones are found.
\begin{CodeInput}
genClones(object, chr, tol = 0.9, id = "Genotype")
fixClones(object, gc, id = "Genotype", consensus = TRUE)
\end{CodeInput}
The \code{genClones()} function uses the power of \code{comparegeno()}
from the \pkg{qtl} package to perform the relatedness
calculations. It then provides a numerical breakdown of the
relatedness between pairs of individuals that share a proportion of
alleles greater than \code{tol}. This breakdown also includes the
clonal group the pairs of individuals belong to. The table of
information from this calculation can then be passed to
\code{fixClones()} through the argument \code{gc} and consensus
genotypes are formed through the appropriate merging of alleles across
genotypes within clone groups.

During the linkage map construction process there may be a requirement to
break or merge linkage groups. \pkg{ASMap} provides two functions to
achieve this.
\begin{CodeInput}
breakCross(cross, split = NULL, suffix = "numeric", sep = ".")
mergeCross(cross, merge = NULL, gap = 5)
\end{CodeInput}
The \code{breakCross()} function allows users to break linkage groups
in a variety ways. The \code{split} argument takes a list with
elements named by the linkage group names that require splitting and
containing the markers that immediately proceed where the splits are to be made.
The \code{mergeCross()} function provides a method for merging linkage
groups. Its argument \code{merge} requires a list with elements named
by the proposed linkage group names required and containing the
linkage groups to be merged. It should be noted that this function
places an artificial genetic distance \code{gap}
between the merged linkage groups. Accurate distance estimation would
require a separate map estimation procedure after merging has taken place.

In \pkg{qtl} genetic distances can be estimated using
\code{est.map()} or through \code{read.cross()} when setting the argument
\code{estimate.map = TRUE}. The estimation uses the multi-locus hidden Markov model
technology of \cite{lg87}. Unfortunately this is computationally cumbersome if there are many
markers on a linkage group and becomes more so if there are many
missing allele calls and genotyping errors present. \pkg{ASMap}
contains a small map estimation function that circumvents this computational burden.
\begin{CodeInput}
quickEst(object, chr, map.function = "kosambi", ...)
\end{CodeInput}
The \code{quickest()} function makes use of another function in \pkg{qtl} called
\code{argmax.geno()}. This function is also a multi-locus hidden
Markov algorithm that uses the observed markers present in a linkage
group to impute pseudo-markers at any chosen cM genetic distance. In this
case, the requirement is for a reconstruction or imputation at the
markers themselves. For the most accurate imputation to occur there
needs to be an estimate of genetic distance in place and this is
calculated by converting recombination fractions to genetic distances
after calling \code{est.rf()}. As a result, the \code{quickEst()}
function lives up to its namesake by providing efficient and
accurate genetic distance calculations for large linkage maps.

The functions \code{pullCross()} and \code{pushCross()} described in
  Section \ref{sec:pp} are used to create and manipulate extra list
  elements \code{"co.located"}, \code{"seg.distortion"} and
\code{"missing"} associated with different marker types. Unfortunately, these
list elements are not recognized by the native \pkg{qtl} functions. If the
function \code{subset.cross()} is used to subset the object to a
reduced number of individuals then the data component of each of these
elements will not be subsetted accordingly. In addition, the
statistics in the table component of the elements
\code{"seg.distortion"} and \code{"missing"} will be incorrect for the
newly subsetted linkage map.
\begin{CodeInput}
subsetCross(cross, chr, ind, ...)
\end{CodeInput}
This subset function contains identical functionality to
\code{subset.cross()} but also ensures the data components
of the extra list elements \code{"co.located"}, \code{"seg.distortion"} and
\code{"missing"} are subsetted to match the linkage map. In addition,
for elements \code{"seg.distortion"} and \code{"missing"} it also
updates the table components to reflect the newly subsetted
map, ensuring \code{pushCross()} uses the most accurate information
when determining which markers to push back into the linkage map.

There is often a requirement to incorporate additional markers to an
established linkage map or merge two linkage maps from the same
population. This idea motivated the creation of the
\code{combineMap()} \pkg{ASMap} package. The aim of the function was
to merge linkage maps based on shared map information, readying the combined
linkage groups for reconstruction through an efficient linkage map
construction process such as \code{mstmap.cross()}.
\begin{CodeInput}
combineMap(..., id = "Genotype", keep.all = TRUE)
\end{CodeInput}
The function takes an unlimited number of linkage maps through the \code{...}
argument. The linkage maps must all have the same cross class
structure and contain the same genotype identifier \code{id}. The merging
of the maps happens intelligently with initial merging based on commonality between
the genotypes. If \code{keep.all = TRUE} the new combined linkage map
is ``padded out'' with missing values where genotypes are not
shared. If \code{keep.all= FALSE} the combined map is reduced to
genotypes that are shared among all linkage maps. Secondly, if linkage
group names are shared between maps then the markers from common
linkage groups are clustered.

\section{Illustrative example}
\label{sec:work}

To showcase the functionality of the \pkg{ASMap} package, the complete
linkage map construction process is presented
for a barley Backcross population containing 3024 markers genotyped
on 326 individuals in an unconstructed marker set formatted as a \pkg{qtl} object with class
\code{"bc"}. The data is available in the \pkg{ASMap} package using
\begin{Schunk}
\begin{Sinput}
R> library("ASMap")
R> data("mapBCu")
\end{Sinput}
\end{Schunk}
\subsection{Pre-construction}

Before constructing a linkage map it is prudent to go through a
pre-construction checklist to ensure that the best quality
genotypes/markers are being used to construct the linkage map. A
non-exhaustive ordered checklist for an unconstructed marker set could
be
\begin{list}{$\bullet$}{\setlength{\itemsep}{1ex}\setlength{\parsep}{0ex}\setlength{\parskip}{0.5ex}}
 \item Check missing allele scores across markers for each genotype
   as well as across genotypes for each marker. Markers or genotypes
   with a high proportion of missing information could be problematic.
 \item Check for genetic clones or individuals that have
   a high proportion of matching allelic information between them.
\item Check markers for excessive segregation distortion. Highly
   distorted markers may not map to unique locations.
 \item Check markers for switched alleles. These markers will not
  cluster or link well with other markers during the construction
  process and it is therefore preferred to repair their alignment before
  proceeding.
 \item Check for co-locating markers. For large linkage maps it would
   be more computationally efficient from a construction standpoint to
   temporarily omit markers that are co-located with other markers.
\end{list}

\begin{figure}[t]
\centering
\includegraphics[width = 16cm]{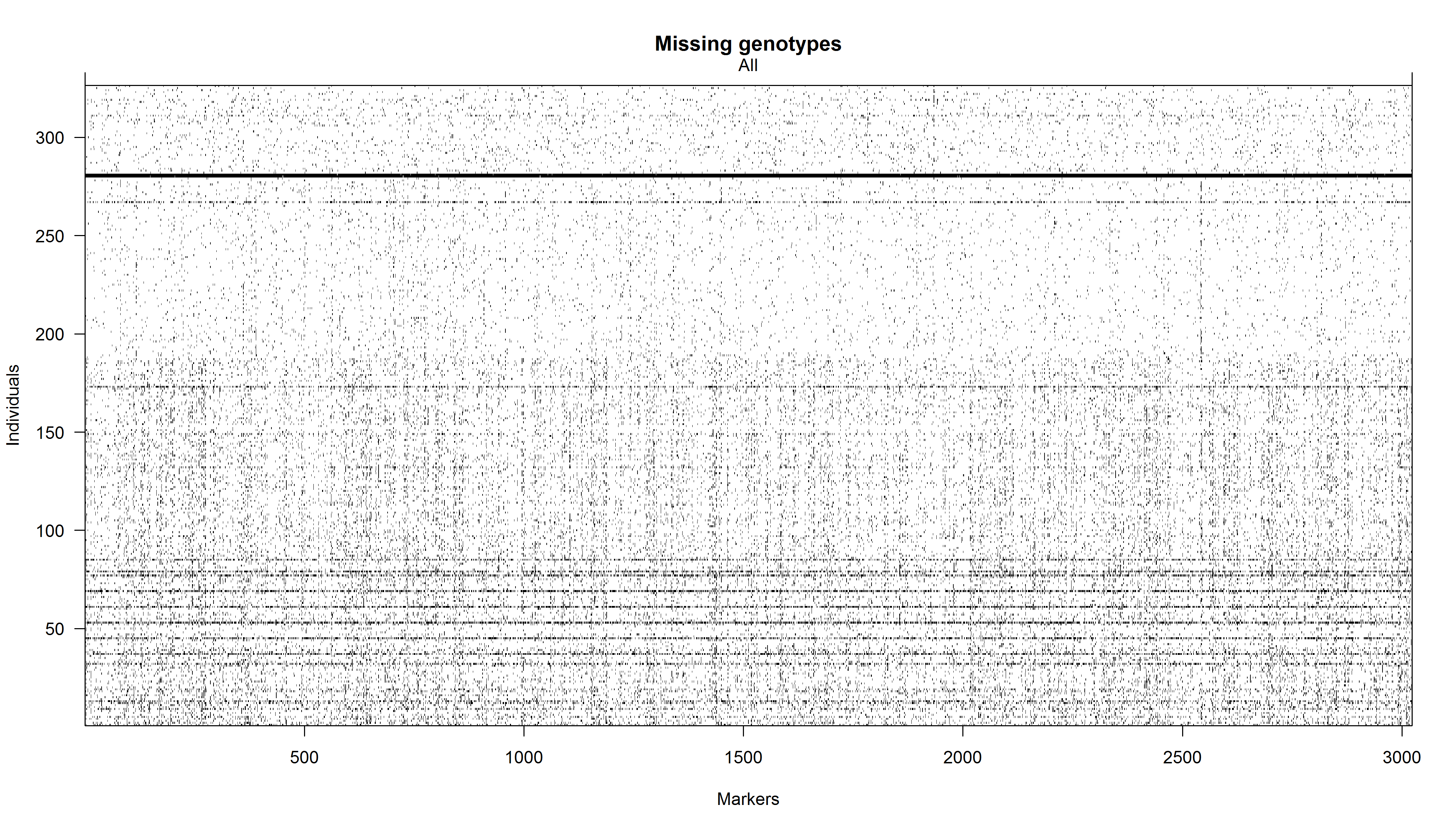}
\caption{Plot of the missing allele scores for the unconstructed map mapBCu.}
\label{fig:miss1}
\end{figure}

\begin{figure}[t]
\includegraphics{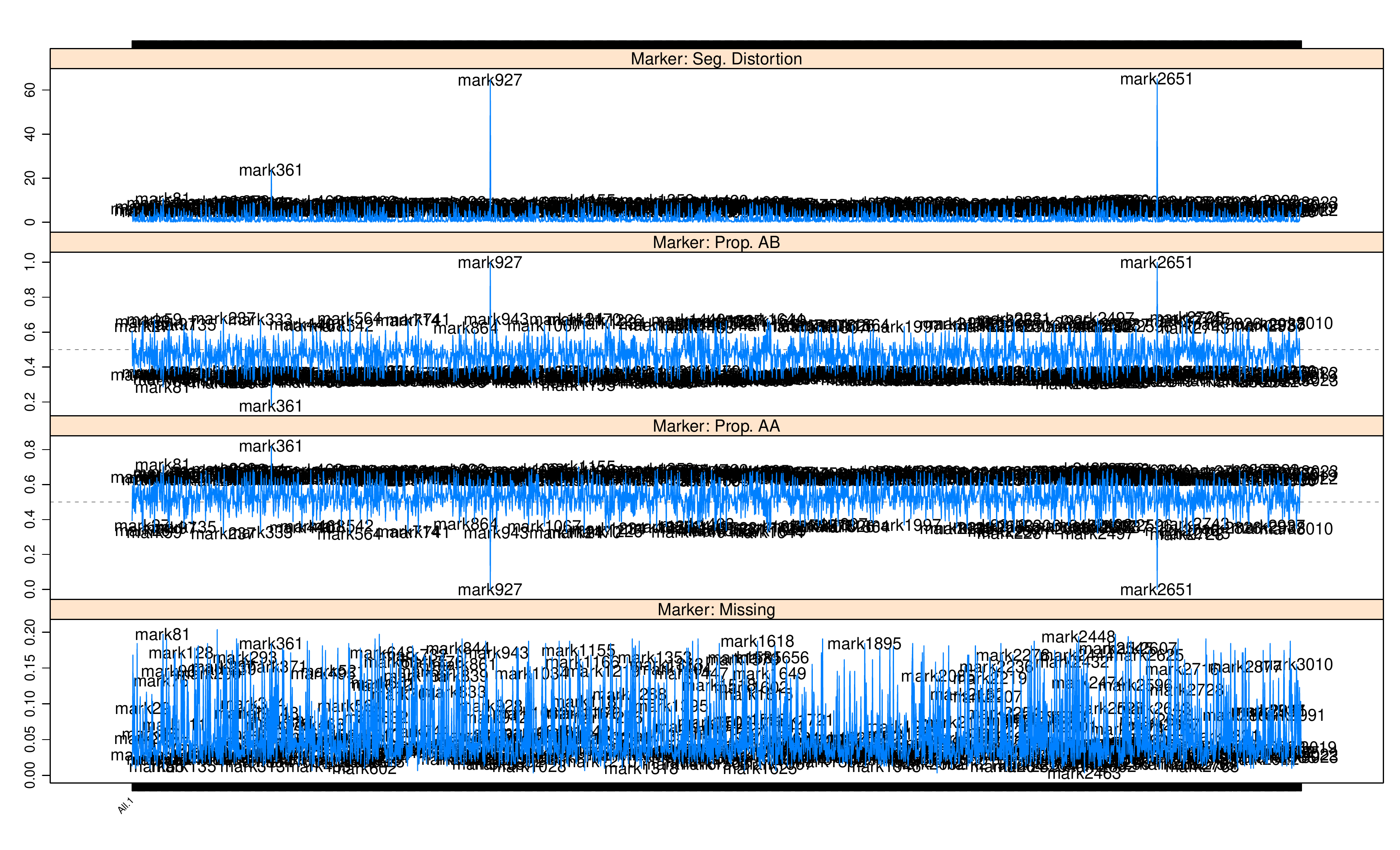}
\caption{For individual markers, the negative log10 p-value for the test of
  segregation distortion, the proportion of each contributing allele
  and the proportion of missing values.}
\label{fig:seg1}
\end{figure}

Figure \ref{fig:miss1} shows the result of a call to the missing
  value diagnostic plot \code{plot.missing()} available in \pkg{qtl}.
\begin{Schunk}
\begin{Sinput}
R> plot.missing(mapBCu)
\end{Sinput}
\end{Schunk}
The darkest horizontal lines on the plot indicate there are some genotypes with large
amounts of missing data. This could indicate poor physical genotyping
of these lines and should be removed before proceeding. The plot
also reveals the markers have a large number of typed allele
values across the range of genotypes. The \pkg{ASMap} function
\code{statGen()} was used to identify genotypes with more than 50\%
missing alleles across the markers. These were omitted using
the usual functions available in \pkg{qtl}.
\begin{Schunk}
\begin{Sinput}
R> sg <- statGen(mapBCu, bychr = FALSE, stat.type = "miss")
R> mapBC1 <- subset(mapBCu, ind = sg$miss < 1600)
\end{Sinput}
\end{Schunk}
From a map construction point of view, highly related individuals may enhance segregation
distortion of markers. It is therefore wise to determine a course of
action such as removal of individuals or the creation of consensus genotypes before proceeding
with any further pre-construction diagnostics. The \pkg{ASMap} function
\code{genClones()} discussed in Section \ref{sec:misc} was used to
identify and report genetic clones.
\begin{Schunk}
\begin{Sinput}
R> gc <- genClones(mapBC1, tol = 0.95)
R> gc$cgd
\end{Sinput}
\begin{Soutput}
      G1    G2   coef match diff na.both na.one group
1  BC045 BC039 0.9919  1466   12      97   1448     1
2  BC052 BC039 1.0000  2423    0      98    502     1
3  BC168 BC039 1.0000  2572    0      47    404     1
4  BC052 BC045 0.9899  1476   15      94   1438     1
5  BC168 BC045 0.9872  1620   21      44   1338     1
6  BC168 BC052 1.0000  2577    0      36    410     1
7  BC067 BC060 1.0000  2759    0       8    256     2
8  BC135 BC086 1.0000  2743    0      17    263     3
9  BC099 BC093 1.0000  2737    0      19    267     4
10 BC120 BC117 1.0000  2699    0      22    302     5
11 BC129 BC126 1.0000  2678    0      35    310     6
12 BC204 BC138 1.0000  2691    0       6    326     7
13 BC147 BC141 0.9996  2753    1      22    247     8
14 BC193 BC144 1.0000  2771    0       7    245     9
15 BC162 BC161 1.0000  2686    0      20    317    10
16 BC205 BC190 1.0000  2886    0       7    130    11
17 BC286 BC285 1.0000  2920    0       1    102    12
18 BC325 BC314 1.0000  2911    0       4    108    13
\end{Soutput}
\end{Schunk}
The table shows 13 groups of genotypes that share a proportion of their
alleles greater than 0.95. The supplied additional statistics show that
the first group contains three pairs of genotypes that had matched
pairs of alleles from 1620 markers or less. These pairs also had $\sim 1400$
markers where an allele was present for one genotype and missing for
another. Based on this, there was not enough evidence to suspect these
pairs may be clones and they were removed from the table. The
\code{fixClones()} function was then be used to form consensus
genotypes for the remaining groups of clones in the table.
\begin{Schunk}
\begin{Sinput}
R> cgd <- gc$cgd[-c(1, 4, 5),]
R> mapBC2 <- fixClones(mapBC1, cgd, consensus = TRUE)
\end{Sinput}
\end{Schunk}
At this juncture it is wise to check whether the observed allelic
frequencies at a specific loci deviate from expected allelic
frequencies. This is called segregation
distortion and it is well known to occur from errors in the physical
laboratorial processes and can also occur in local genomic regions from
underlying biological and genetic mechanisms \citep{lytt91}. The
significance of the segregation distortion, the allelic proportions and
the missing value proportion across the genome can be graphically
represented using the marker profiling function
\code{profileMark()} and the result is displayed in Figure
\ref{fig:seg1}.
\begin{Schunk}
\begin{Sinput}
R> profileMark(mapBC2, stat.type = c("seg.dist", "prop", "miss"), crit.val =
+   "bonf", layout = c(1, 4), type = "l")
\end{Sinput}
\end{Schunk}
Setting \code{crit.val = "bonf"} annotates the markers
in each panel that have a p-value for the test of segregation distortion lower than
the family wide bonferroni adjusted alpha level of 0.05/(total number
of markers). The plot indicates there are numerous markers that
are significantly distorted with three highly distorted
markers. The plot also shows the missing value proportion of
the markers does not exceed 20\%.

The highly distorted markers were omitted using
\begin{Schunk}
\begin{Sinput}
R> mm <- statMark(mapBC2, stat.type = "marker")$marker$AB
R> dm <- markernames(mapBC2)[(mm > 0.98) | (mm < 0.2)]
R> mapBC3 <- drop.markers(mapBC2, dm)
\end{Sinput}
\end{Schunk}

Without a constructed map it is impossible to determine the origin of
the segregation distortion. However, the blind use of distorted markers may also
create linkage map construction problems. It may
be more sensible to place the distorted markers aside and construct
the map with less problematic markers. Once the linkage map is constructed the
more problematic markers can be introduced to determine whether they
have a useful or deleterious effect on the map. The \pkg{ASMap} functions
\code{pullCross()} and \code{pushCross()} discussed in Section
\ref{sec:pp} are designed to take advantage of this scenario. To
showcase their use in this example, they were used to pull markers
with 10-20\% missing values, markers with significant segregation
distortion and co-located markers.

\begin{Schunk}
\begin{Sinput}
R> pp <- pp.init(miss.thresh = 0.1, seg.thresh = "bonf")
R> mapBC3 <- pullCross(mapBC3, type = "missing", pars = pp)
R> mapBC3 <- pullCross(mapBC3, type = "seg.distortion", pars = pp)
R> mapBC3 <- pullCross(mapBC3, type = "co.located")
R> names(mapBC3)
\end{Sinput}
\begin{Soutput}
[1] "geno"           "pheno"          "missing"       
[4] "seg.distortion" "co.located"    
\end{Soutput}
\end{Schunk}
A total of 847 markers were removed and placed aside in their
respective elements, with 2173 markers remaining in the map ready for
linkage map construction.

\begin{figure}[t]
 \centering
\includegraphics[width = 16cm]{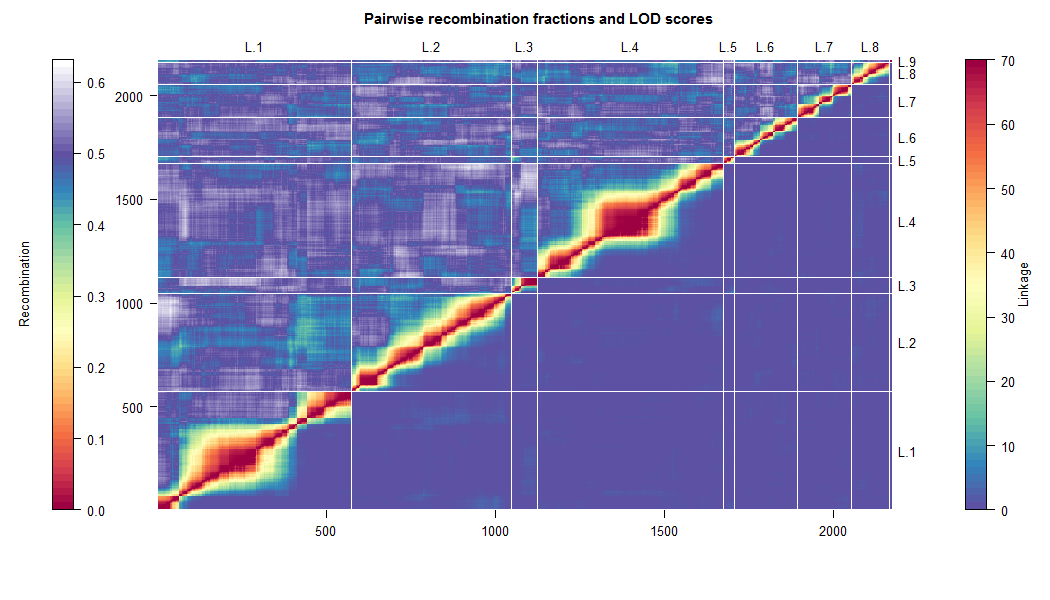}
\caption{Heat map of the constructed linkage map mapBC4.}
\label{fig:heat3}
\end{figure}

\subsection{MSTmap construction}

The curated genetic marker data in \code{mapBC3} was then constructed using the
\code{mstmap.cross()} function.

\begin{Schunk}
\begin{Sinput}
R> mapBC4 <- mstmap(mapBC3, bychr = FALSE, trace = TRUE, p.value = 1e-12)
R> chrlen(mapBC4)
\end{Sinput}
\begin{Soutput}
       L.1        L.2        L.3        L.4        L.5 
304.910957 266.240647  78.982131 252.281760  33.226962 
       L.6        L.7        L.8        L.9 
233.485952 153.315888 106.290403   6.657526 
\end{Soutput}
\end{Schunk}
By setting \code{bychr = FALSE} the complete set of marker data from
\code{mapBC3} is bulked and constructed from scratch. This
construction involves the clustering of markers to linkage groups and
the optimal ordering of markers within each linkage group. Figure
\ref{fig:threshold}, indicates for
a population size of 309 the \code{p.value} should be set to $10^{-12}$
to ensure a 30cM threshold when clustering markers to linkage groups.
The newly constructed linkage map contains nine linkage groups each
containing markers that are optimally ordered. The performance of the
MSTmap construction was checked by plotting the heat map of
pairwise recombination fractions (RFs) between markers and their pairwise
LOD score of linkage using
\begin{Schunk}
\begin{Sinput}
R> heatMap(mapBC4, lmax = 70)
\end{Sinput}
\end{Schunk}
and is given in Figure \ref{fig:heat3}. An aesthetic heat map is
attained when the heat on the lower triangle of the plot (pairwise LOD scores)
matches the heat on the upper triangle (pairwise estimated RFs) and
this was achieved by setting \code{lmax = 70}. The heat map shows
consistent heat across the markers within linkage
groups indicating strong linkage between nearby markers. The linkage
groups appear to be very distinctly clustered.

Although the heat map is indicating the construction process was
successful it does not highlight subtle problems that may be existing
in the constructed linkage map. One of the key quality characteristics
of a well constructed linkage map is an appropriate recombination rate
of each of the genotypes. For a conservative theorised chromosomal
length of 200cM each member of the progeny of the barley backcross
population line have an approximate expected recombination rate of ~ 14 across the
genome. Genotypes that significantly exceeed this rate were checked
using the \code{profileGen()} function.

\begin{figure}[t]
\includegraphics{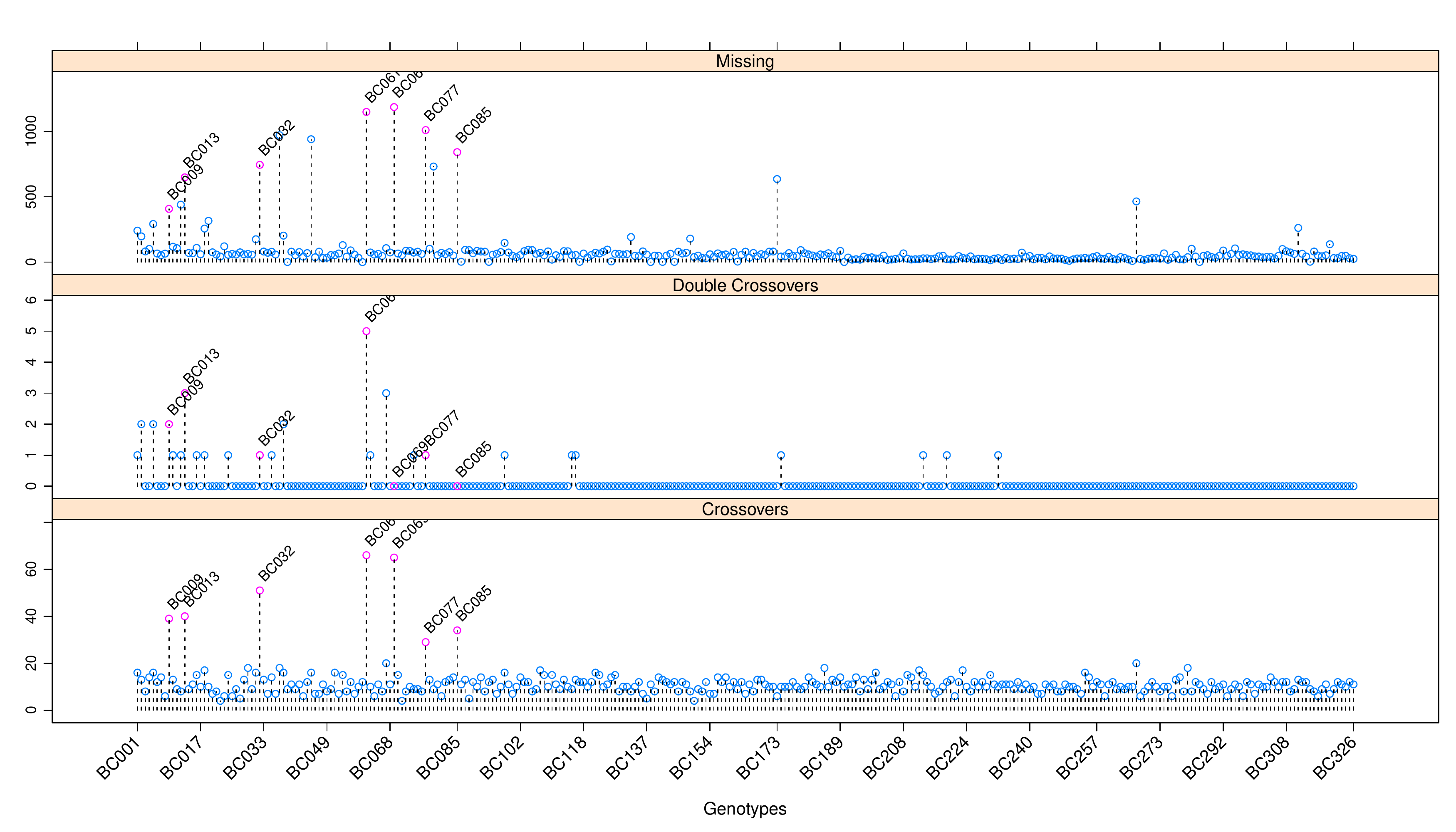}
\caption{For individual genotypes, the number of recombinations,
  double recombinations and missing values for mapBC4.}
\label{fig:genex1}
\end{figure}

\begin{Schunk}
\begin{Sinput}
R> pg <- profileGen(mapBC4, bychr = FALSE, stat.type = c("xo", "dxo",
+   "miss"), id = "Genotype", xo.lambda = 14, layout = c(1, 3), lty = 2)
\end{Sinput}
\end{Schunk}

Figure \ref{fig:genex1} show the number of recombinations, double recombination
and missing values for each of 309 genotypes. The plot also annotates
the genotypes that have recombination rates significantly above an expected
recombination rate of 14. A total of seven lines have recombination
rates above 20 and the plots also show that these lines have
excessive missing values. To ensure the extra list elements
\code{"co.located"}, \code{"seg.distortion"} and
\code{"missing"} of the object are subsetted and updated appropriately, the offending
genotypes were removed using the \pkg{ASMap} function
\code{subsetCross()}. The linkage map was then reconstructed.
\begin{Schunk}
\begin{Sinput}
R> mapBC5 <- subsetCross(mapBC4, ind = !pg$xo.lambda)
R> mapBC6 <- mstmap(mapBC5, bychr = TRUE, trace = TRUE, p.value = 1e-12)
R> chrlen(mapBC6)
\end{Sinput}
\begin{Soutput}
       L.1        L.2        L.3        L.4        L.5 
229.578103 223.170605  65.806120 214.380402  27.297642 
       L.6        L.7        L.8        L.9 
191.060846 140.114968  88.687053   4.875885 
\end{Soutput}
\end{Schunk}
As a result the lengths of the linkage groups dropped dramatically.

\begin{figure}[t]
\includegraphics{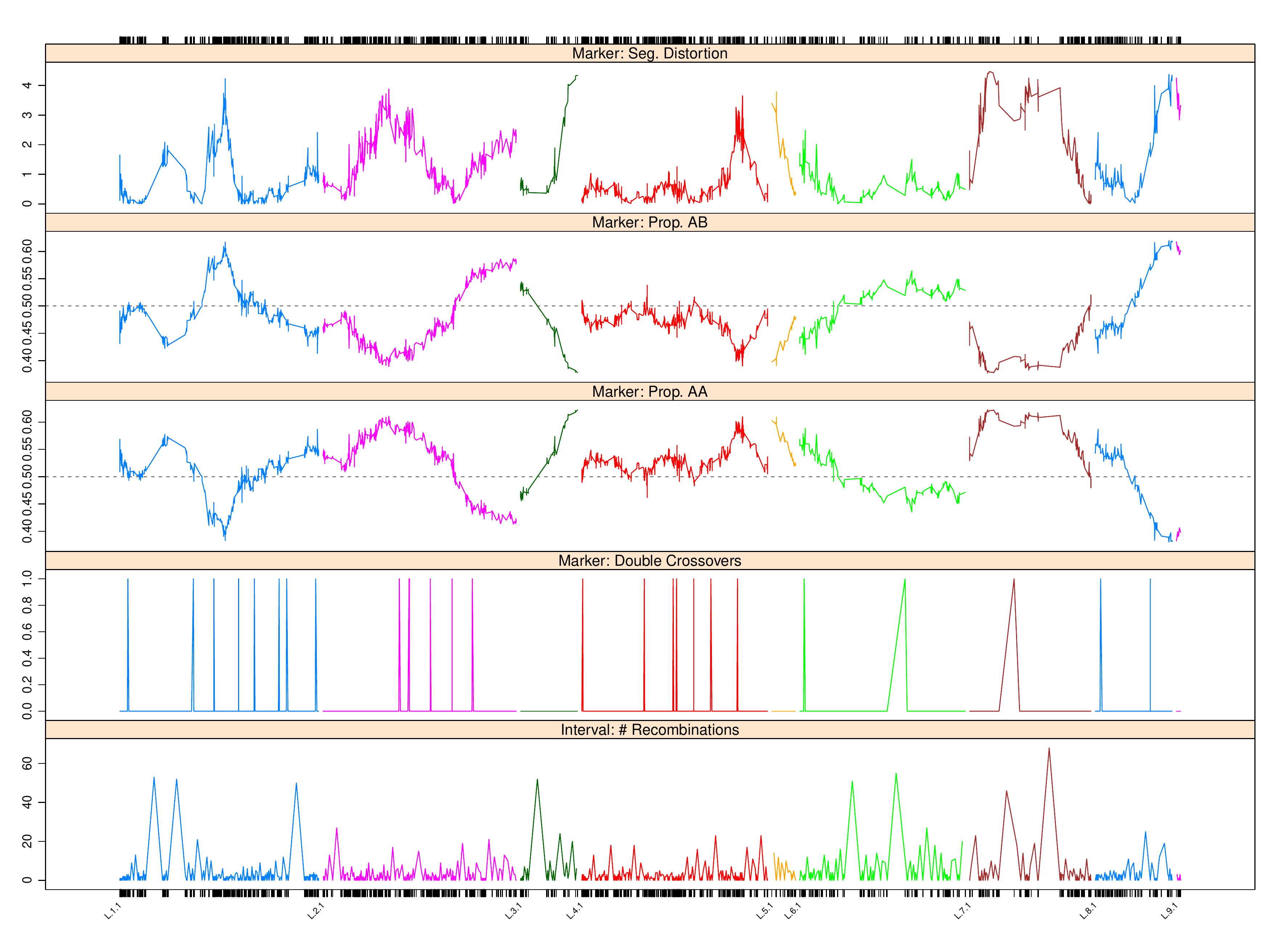}
\caption{Marker profiles of the negative log10 p-value for the test of
  segregation distortion, allele proportions and the number of double
  crossovers as well as the interval profile of the number of
  recombinations between adjacent markers in mapBC6.}
\label{fig:pm1}
\end{figure}

It is also useful to graphically display statistics of
the markers and intervals of the current constructed linkage map. For
example, Figure \ref{fig:pm1} shows the marker profiles of the -log10 p-value
for the test of segregation distortion, the allele proportions and the
number of double crossovers. It also displays the interval profile of
the number of recombinations occurring between adjacent markers. This
plot reveals many things that are useful for the next phase of the
construction process.
\begin{Schunk}
\begin{Sinput}
R> profileMark(mapBC6, stat.type = c("seg.dist", "prop", "dxo", "recomb"),
+   layout = c(1, 5), type = "l")
\end{Sinput}
\end{Schunk}

Figure \ref{fig:pm1} reveals the success of the map construction process
with no more than one double crossover being found at any marker and
very few being found in total. The plot also reveals the extent of the
biological distortion that can occur within a linkage group. A close look at the segregation
distortion and allele proportion plots shows the linkage group
L.3 and the short linkage group L.5 have profiles that could be joined
if the linkage groups were merged. In addition, L.8 and L.9 also have
profiles that could be joined if the linkage groups were
combined.

\subsection{Pushing back markers}
\label{sec:push}

\begin{figure}[t]
\includegraphics{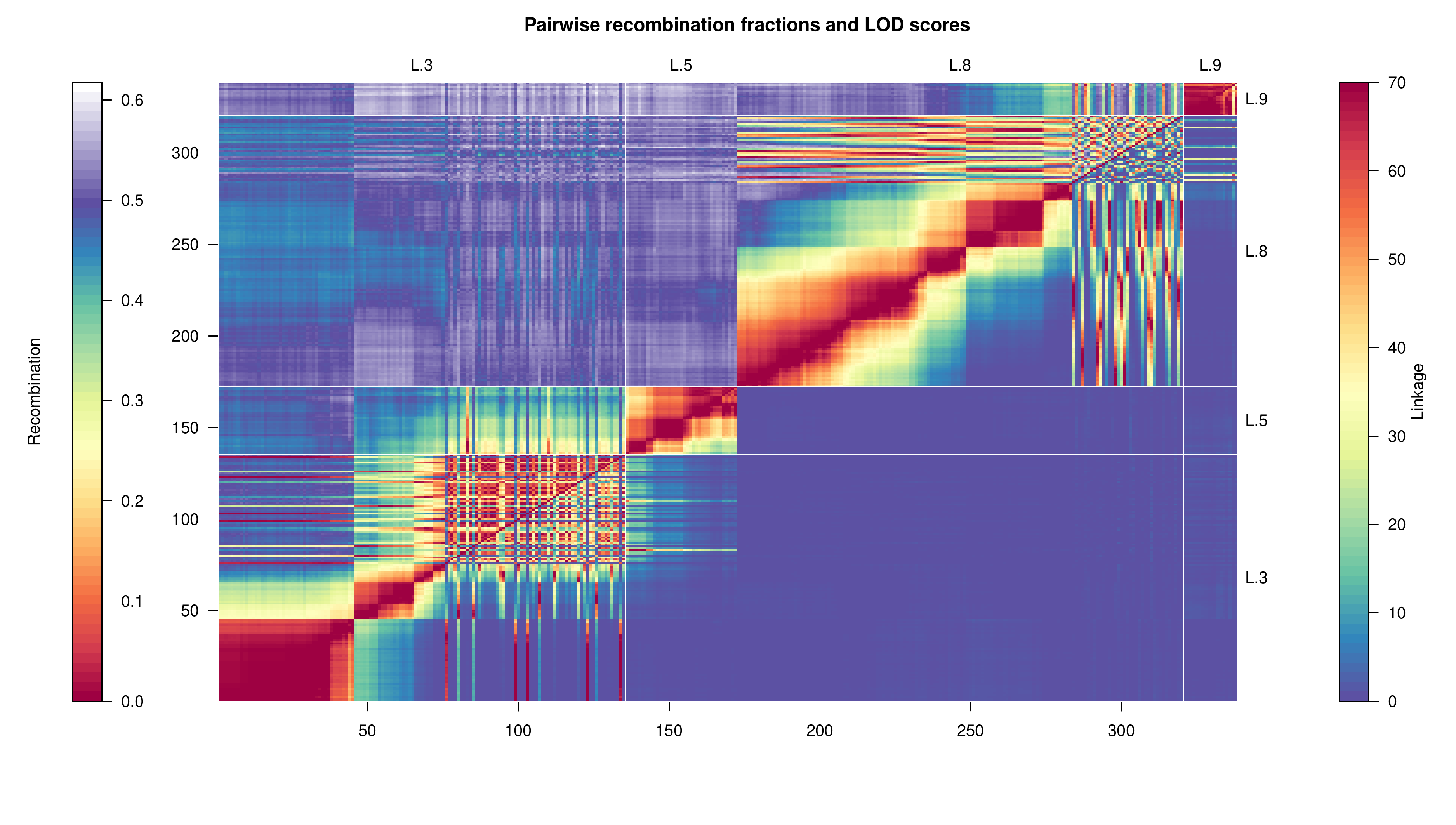}
\caption{Heat map of the linkage map mapBC6 subsetted to linkage
  groups L.3, L.5, L.8 and L.9.}
\label{fig:heat4}
\end{figure}

Markers that were originally placed aside in the
pre-construction of the linkage map can now be pushed back into the
constructed linkage map and the map carefully re-diagnosed. To begin,
the 515 external markers that have between 10\% and 20\% missing
values residing in the list element \code{"missing"} were pushed
back in using \code{pushCross()}
\begin{Schunk}
\begin{Sinput}
R> pp <- pp.init(miss.thresh = 0.22, max.rf = 0.3)
R> mapBC6 <- pushCross(mapBC6, type = "missing", pars = pp)
\end{Sinput}
\end{Schunk}
The parameter \code{miss.thresh = 0.22} was used to ensure that all
markers with a missing value proportion less 0.22 are pushed back into the linkage
map. Note, this pushing mechanism is not re-constructing the map and
only assigns the markers to the most suitable linkage group. At this
point it is worth re-plotting the heat map to check whether the push
has been successful. The graphic is confined to linkage groups L.3, L.5,
L.8 and L.9 to determine whether the extra markers have provided
useful additional information about the possible
merging of the groups. The resulting heat map is given in Figure \ref{fig:heat4}.
\begin{Schunk}
\begin{Sinput}
R> heatMap(mapBC6, chr = c("L.3", "L.5", "L.8", "L.9"), lmax = 70)
\end{Sinput}
\end{Schunk}

It is clear from the heat map that there are genuine linkages between
L.3 and L.5 as well as L.8 and L.9. These two sets of linkage groups
were merged using \code{mergeCross()} and the linkage group names
are renamed to form the optimal seven linkage groups that are required for
the barley genome.

\begin{Schunk}
\begin{Sinput}
R> mlist <- list("L.3" = c("L.3", "L.5"), "L.8" = c("L.8", "L.9"))
R> mapBC6 <- mergeCross(mapBC6, merge = mlist)
R> names(mapBC6$geno) <- paste("L.", 1:7, sep = "")
R> mapBC7 <- mstmap(mapBC6, bychr = TRUE, trace = TRUE, p.value = 2)
R> chrlen(mapBC7)
\end{Sinput}
\begin{Soutput}
     L.1      L.2      L.3      L.4      L.5      L.6      L.7 
242.4104 233.1945 162.6205 224.1773 201.1503 147.7669 159.5308 
\end{Soutput}
\end{Schunk}

As the optimal number of linkage groups have been identified the
re-construction of the map was performed by linkage group only
through setting \code{p.value = 2}. The linkage group lengths of L.1,
L.2 and L.4 are slightly elevated indicating excessive recombination
across these groups.
\begin{Schunk}
\begin{Sinput}
R> pg1 <- profileGen(mapBC7, bychr = FALSE, stat.type = c("xo", "dxo",
+   "miss"), id = "Genotype", xo.lambda = 14, layout = c(1,3), lty = 2)
\end{Sinput}
\end{Schunk}

\begin{figure}[t]
\includegraphics{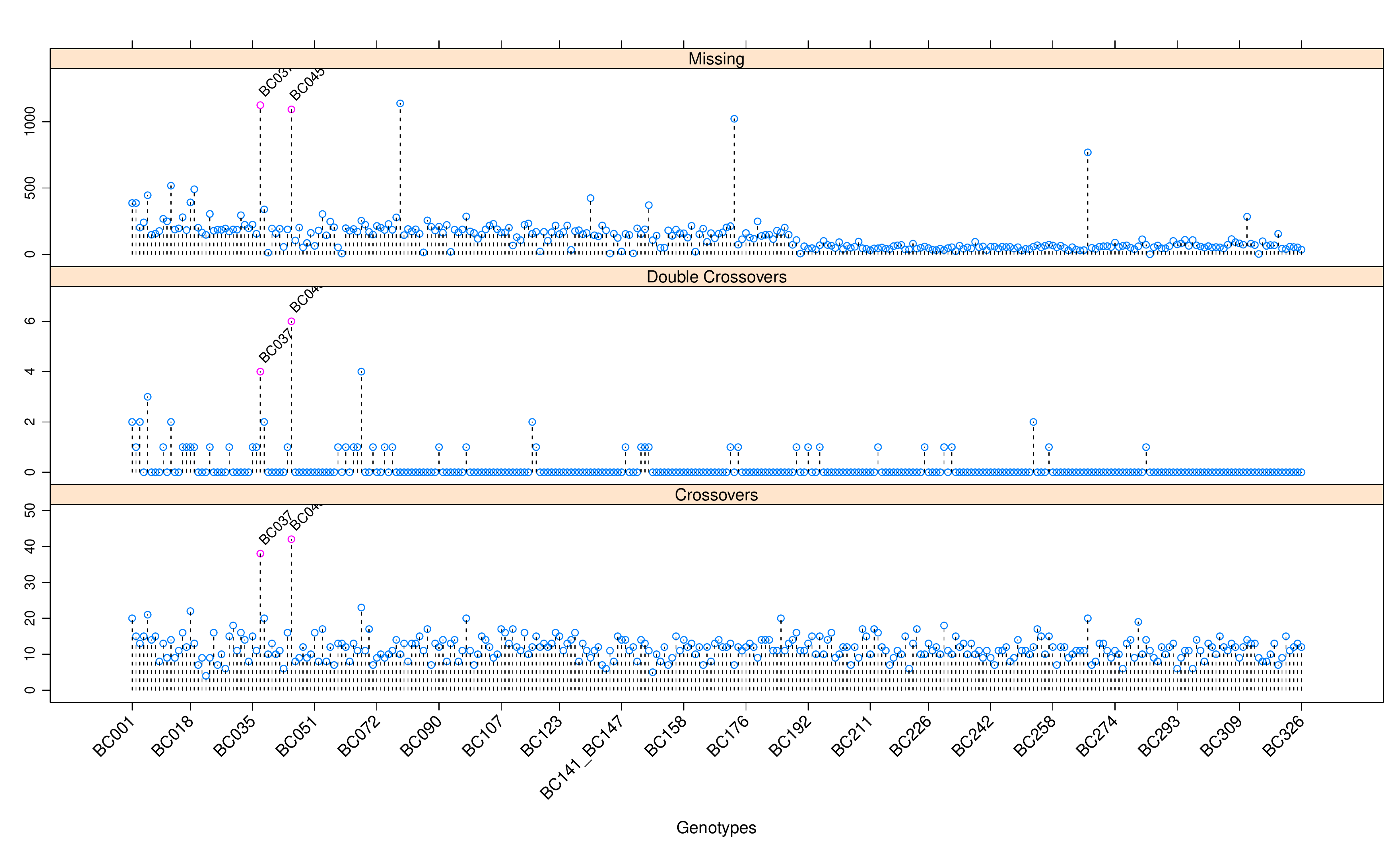}
\caption{For individual genotypes, the number of recombinations,
  double recombinations and missing values for mapBC7.}
\label{fig:genex2}
\end{figure}

\begin{figure}[t]
\includegraphics{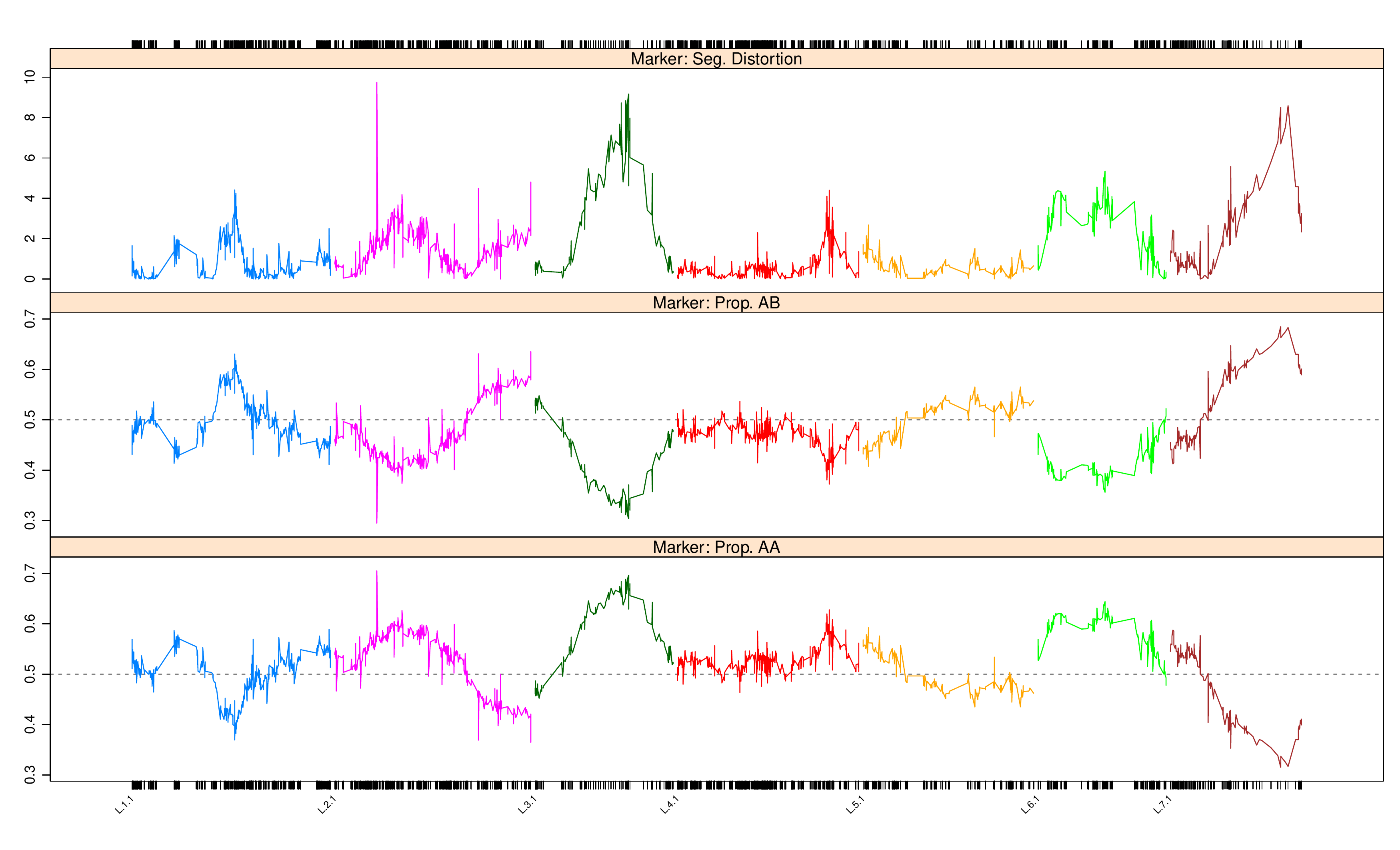}
\caption{Marker profiles of the negative log10 p-value for the test of
  segregation distortion, allele proportions for mapBC9}
\label{fig:pm2}
\end{figure}

Figure \ref{fig:genex2} shows the genotype profiles for the 302 barley
lines. The introduction of the markers with missing value proportions between
10\% and 20\% into the linkage map has highlighted two more
problematic lines that have a high proportion of missing values across
the genome. These lines were removed using
\code{subsetCross()} and the map reconstructed.
\begin{Schunk}
\begin{Sinput}
R> mapBC8 <- subsetCross(mapBC7, ind = !pg1$xo.lambda)
R> mapBC9 <- mstmap(mapBC8, bychr = TRUE, trace = TRUE, p.value = 2)
R> chrlen(mapBC9)
\end{Sinput}
\begin{Soutput}
     L.1      L.2      L.3      L.4      L.5      L.6      L.7 
225.7900 223.3063 157.0592 206.4801 194.3003 145.7554 149.2467 
\end{Soutput}
\end{Schunk}
The removal of these two lines will not have a deleterious effect on the number
of linkage groups and therefore the linkage map was
reconstructed by linkage group only. The length of most linkage groups
has now been appreciably reduced. Figure \ref{fig:pm2} displays the segregation distortion
and allele proportion profiles for the markers from using \code{profileMark()} again.
\begin{Schunk}
\begin{Sinput}
R> profileMark(mapBC9, stat.type = c("seg.dist", "prop"), layout = c(1, 5),
+   type = "l")
\end{Sinput}
\end{Schunk}

The plot indicates a spike of segregation distortion on L.2 that does
not appear to be biological and needs to be removed. The plot also
indicates the significant distortion regions on L.3, L.6 and
L.7 indicating some of the markers with missing value proportions
between 10\% and 20\% pushed back into the linkage map also had some
degree of segregation distortion. The marker positions on the x-axis of the
plot suggests these regions are sparse. The 295 external markers in the
list element \code{"seg.distortion"} may hold the key to this
sparsity and were pushed back to determine their effect on the linkage
map.
\begin{Schunk}
\begin{Sinput}
R> sm <- statMark(mapBC9, chr = "L.2", stat.type = "marker")
R> dm <- markernames(mapBC9, "L.2")[sm$marker$neglog10P > 6]
R> mapBC10 <- drop.markers(mapBC9, dm)
R> pp <- pp.init(seg.ratio = "70:30")
R> mapBC11 <- pushCross(mapBC10, type = "seg.distortion", pars = pp)
R> mapBC12 <- mstmap(mapBC11, bychr = TRUE, trace = TRUE, p.value = 2)
R> round(chrlen(mapBC12) - chrlen(mapBC9), 5)
\end{Sinput}
\begin{Soutput}
     L.1      L.2      L.3      L.4      L.5      L.6      L.7 
 0.20398  1.11431  0.60228 -0.92150  0.86448 -0.25043 -6.29943 
\end{Soutput}
\begin{Sinput}
R> nmar(mapBC12) - nmar(mapBC10)
\end{Sinput}
\begin{Soutput}
L.1 L.2 L.3 L.4 L.5 L.6 L.7 
  0   1 156   0   0  86  52 
\end{Soutput}
\end{Schunk}
After checking the table element of the \code{"seg.distortion"}
element, a 70:30 distortion ratio was chosen to ensure all distorted
markers were pushed back into the linkage map. After the pushing was
complete, the linkage map was reconstructed and linkage group lengths
changed negligibly from the previous version of the map.
This indicates the extra markers have been inserted
successfully with nearly all distorted markers being pushed into
L.3, L.6 and L.7.

To finalise the map the 39 co-locating markers residing in the
\code{"co.located"} list element of the object were pushed back into
the linkage map and placed adjacent to the markers they were co-located
with. Note that all external co-located markers have an immediate linkage group assignation.
\begin{Schunk}
\begin{Sinput}
R> mapBC <- pushCross(mapBC12, type = "co.located")
R> names(mapBC)
\end{Sinput}
\begin{Soutput}
[1] "geno"  "pheno"
\end{Soutput}
\end{Schunk}
A check of the final structure of the object shows the extra
list elements have been removed and only the \code{"pheno"}
and \code{"geno"} list elements remain. The final linkage map
statistics are given in Table \ref{tab:stats}.

\begin{table}[t]
\centering
\begin{tabular}{lrrrrrrrrr}
  \hline
 & L.1 & L.2 & L.3 & L.4 & L.5 & L.6 & L.7 & Total & Ave. \\
  \hline
  \hline
No. of markers & 681 & 593 & 335 & 679 & 233 & 279 & 219 & 3019 & 431 \\
Lengths & 225.9 & 224.4 & 157.6 & 205.5 & 195.1 & 145.5 & 142.9 & 1297.2 & 185.3 \\
Ave. interval & 0.33 & 0.39 & 0.48 & 0.31 & 0.84 & 0.53 & 0.66 &  & 0.51 \\
   \hline
\end{tabular}
\caption{Table of statistics for the final linkage map, mapBC.}
\label{tab:stats}
\end{table}

\subsection{Post construction linkage map development}

\pkg{ASMap} provides functionality to insert additional markers
into an established linkage map without losing important
linkage group identification. The methods applied in this section
assume the additional markers, as well as the constructed linkage
map, are \pkg{qtl} cross objects of the same class. The methods are best described by
presenting several examples that mimic common post-construction
linkage map development tasks. In these examples
additional markers will be obtained by randomly selecting markers
from the final linkage map obtained in the previous
Section, \code{mapBC}.
\begin{Schunk}
\begin{Sinput}
R> set.seed(123)
R> mn <- markernames(mapBC)[sample(1:3019, 2700, replace = FALSE)]
R> add1 <- drop.markers(mapBC, mn)
R> mapBCs <- drop.markers(mapBC, markernames(add1))
R> add2 <- subset(add1, chr = "L.1")
R> add3 <- add1
R> mn3 <- markernames(add3)
R> for(i in 1:length(mn3))
+       add3 <- movemarker(add3, mn3[i], "ALL")
\end{Sinput}
\end{Schunk}

\textbf{Combining linkage maps:} Many populations that are currently being researched have been
genotyped on multiple platforms and separate linkage maps constructed for one
or both of the populations. The \code{add1} \pkg{qtl} object mimics an
older map of \code{mapBC} with 319 markers spanning the seven linkage
groups. As the linkage groups are known between maps there is only a
requirement to combine the maps in a sensible manner and
reconstruct. This can be done efficiently using \code{combineMap()}
without external manipulation and loss of linkage group information.
\begin{Schunk}
\begin{Sinput}
R> add1 <- subset(add1, ind = 2:300)
R> full1 <- combineMap(mapBCs, add1, keep.all = TRUE)
R> full1 <- mstmap(full1, bychr = TRUE, trace = TRUE, anchor = TRUE,
+   p.value = 2)
\end{Sinput}
\end{Schunk}
The \code{combineMap()} function merged the two maps on their matching
  genotypes with missing values added in the first genotype for the markers from
\code{add1}. The function also understands that
both linkage maps share common linkage group names and placed the
markers from shared linkage groups together. The map can then be
reconstructed by linkage group using \code{p.value = 2} in
the \code{mstmap.cross()} call, ensuring the important
identity of linkage groups are retained. In addition, setting \code{anchor = TRUE}
ensures that the orientation of the linkage groups will be preserved
for the larger linkage map, \code{mapBCs}.

\textbf{Fine mapping:} In marker assisted selection breeding programmes it is common to
increase the density of markers in a specific genomic region of a
linkage group for the purpose of more accurately identifying the
position of gene related loci. This technique is known as \textit{fine
mapping}. The \code{add2} object contains only
markers from the L.1 linkage group of \code{mapBCs}. When the linkage
group for the additional markers is known in advance and matches a
linkage group in the constructed map, the insertion of
the new markers is very similar to the previous section.
\begin{Schunk}
\begin{Sinput}
R> add2 <- subset(add2, ind = 2:300)
R> full2 <- combineMap(mapBCs, add2, keep.all = TRUE)
R> full2 <- mstmap(full2, chr = "L.1", bychr = TRUE, trace = TRUE,
+   anchor = TRUE, p.value = 2)
\end{Sinput}
\end{Schunk}
Again, the removal of the first genotype of \code{add2} causes
missing values to be added in the first genotype of \code{full2} for
the markers that were in \code{add2}. After the maps are combined, all
markers from L.1 are clustered together and only L.1 requires
reconstructing.

\textbf{Unknown linkage groups:} There may be occasions when the
linkage group identification of the
additional markers is not known in advance. For example, an incomplete
set of markers was used to construct the map or a secondary set of
markers are available that come from an unconstructed linkage map.
The \pkg{qtl} object \code{add3} consists of one linkage group
called \code{ALL} containing 319 markers spanning the seven
linkage groups in \code{add1}.
\begin{Schunk}
\begin{Sinput}
R> add3 <- subset(add3, ind = 2:300)
R> full3 <- combineMap(mapBCs, add3, keep.all = TRUE)
R> full3 <- pushCross(full3, type = "unlinked", unlinked.chr = "ALL")
R> full3 <- mstmap(full3, bychr = TRUE, trace = TRUE, anchor = TRUE,
+   p.value = 2)
\end{Sinput}
\end{Schunk}
Again, \code{combineMap()} was used to initially merge linkage maps.
The function \code{pushCross()} was then used to
push the additional markers into the constructed linkage map. By
choosing the marker type argument \code{type = "unlinked"} and
providing the \code{unlinked.chr = "ALL"} the function recognises that
the markers require pushing back into the remaining linkage groups of
\code{full3}. Again, the linkage map reconstruction only required
optimal ordering of the markers within linkage groups.

\section[Limitations of MSTmap and ASMap]{Performance of MSTmap and \pkg{ASMap}}
\label{sec:lim}

\cite{mst08} contains extensive information on the comparative
performance of the MSTmap algorithm for constructing linkage
maps. However, it was not outlined in \cite{mst08} how well
the algorithm scaled for complete linkage map construction of large genetic
marker sets. Some insight can be gained from \cite{ras13} where a
direct comparison of Lep-MAP with the MSTmap algorithm is presented
and indicates efficient results are achievable with 10,000 markers
genotyped on 200 individuals. To showcase the efficiency and
scalability of the MSTmap algorithm in \pkg{ASMap},
Table \ref{tab:time} presents the computational timings for linkage
map construction of varied simulated DH and F2 populations using
\code{mstmap.cross()}. The simulated data sets comprised of all
combinations of (100, 200, 300) individuals and (1K, 2K, 5K, 10K,
20K) markers evenly distributed in five chromosomes. For example, a 20K
marker set consists of five chromosomes each containing 4K
markers. Timings are presented for complete linkage map
construction (marker clustering and optimal ordering within linkage
groups) as well as marker ordering only within established linkage
groups. All timings are averaged over the five chromosomes.
For brevity, the simulations for DH marker data sets have been restricted
to two error rates, no error rate and one that is comparable to a
realistic downstream error rate obtained from a Genotype by Sequencing
(GBS) pipeline (0.42\%) discussed in \cite{glaub14}. For the latter,
the argument \code{detectBadData = TRUE} has been set in
\code{mstmap.cross()}. F2 marker data sets have only been simulated
with no error due to the lack of error detection capabilities of
the MSTmap algorithm for non-Advanced RIL populations. All simulations
were performed on a Linux Ubuntu 14.04 box with a quad-core 4.7 Ghz
Pentium i7 with 16Gb RAM.

\setlength{\tabcolsep}{4.9pt}
\begin{table}[ht]
\centering
{\small
\begin{tabular}{llllllllllllll}
  \hline
  &  &  & \multicolumn{5}{l}{Clustering and ordering} & &
 \multicolumn{5}{l}{Ordering only}\\
 \cline{4-8} \cline{10-14}
\T & $n$ & & 1K & 2K & 5K & 10K & 20K & & 1K & 2K & 5K & 10K & 20K \\
  \hline
  \hline
 \T DH  & 100 & & 0.31 & 1.28 & 8.03 & 39.16 & 175.2 & & 0.10 & 0.28 & 1.45 & 6.60 & 31.25 \\
(no error) & 200 & & 0.99 & 3.28 & 22.25 & 85.00 & 341.6 & & 0.18 & 0.60 & 3.08 & 14.29 & 63.27 \\
 & 300 & & 0.98 & 4.95 & 32.38 & 122.2 & 491.1 & & 0.29 & 0.91 & 5.22 & 24.27 & 107.4 \\
\hline
\T DH  & 100 & & 0.41 & 1.44 & 12.55 & 53.21 & 245.3 & & 0.13 & 0.40 & 2.17 & 11.75 & 179.6 \\
(0.42\% error) &  200 & & 0.76 & 3.69 & 26.56 & 117.9 & 579.9 & & 0.27 & 1.31 & 6.93 & 33.73 & 185.0 \\
 & 300 & & 1.51 & 5.74 & 39.57 & 164.7 & 680.7 & & 0.42 & 1.38 & 8.43 & 40.15 & 338.6 \\
\hline
\T F2 & 100 & & 1.26 & 4.86 & 30.86 & 149.2 & 670.0 & & 0.39 & 1.29 & 6.61 & 26.67 & 146.2 \\
 & 200 & & 2.52 & 9.65 & 79.06 & 354.7 & 1304.4 & & 0.84 & 3.33 & 21.51 & 79.60 & 292.1 \\
 & 300 & & 4.14 & 18.20 & 111.2 & 529.7 & 2181.2 & & 1.75 & 6.60 & 34.35 & 142.6 & 385.9 \\
   \hline
\end{tabular}}
\caption{Computational timings of MSTmap in \pkg{ASMap} (in seconds) for various simulated DH
  and F2 populations with all combinations of (100, 200, 300)
  individuals and (1K, 2K, 5K, 10K, 20K) markers distributed
  evenly across five chromosomes. Timings are averaged over
  the five chromosomes.}
\label{tab:time}
\end{table}

Table \ref{tab:time} indicates very efficient results are achievable for complete
linkage map construction across the range of marker sets and
population sizes. Without the requirement for initial clustering of markers,
there is a dramatic reduction in computation time required for
optimally ordering of markers within established linkage
groups. Compared to its error free counterpart, there is a moderate
increase in computational time for linkage map construction of marker
sets containing error (0.42\%) caused by the requirement to identify
and remedy potential genotyping errors as the algorithm
progresses. For simulated F2 marker sets, the reduction in
computational efficiency is due to the iterative optimization procedure
required to estimate pairwise information between the markers.

There is clear performance disadvantage when an initial clustering of
markers is included in the MSTmap algorithm. This was also indicated
in \cite{ras13} and \cite{bbc14} where MSTmap had memory limitation
problems attempting to cluster >
100K markers genotyped across a small set of individuals. These
performance issues are due to the potentially enormous number of
pairwise comparisons that require calculation before clustering markers
to linkage groups. However, this computational burden can almost be
completely removed by exploiting the knowledge that, for small
population sizes, there are large numbers of co-locating markers
existing throughout the unconstructed marker set. Consider a simulated
data set of over 50K markers genotyped on 300 individuals created from
the final linkage map \code{mapBC} constructed in Section
\ref{sec:push}.
\begin{Schunk}
\begin{Sinput}
R> simBC <- sim.geno(mapBC, step = 0.025, n.draws = 1, map.function =
+   "kosambi")
R> simBC$geno[["ALL"]]$data <- pull.draws(simBC)[,,1]
R> simBC$geno[["ALL"]]$map <- 1:ncol(simBC$geno[["ALL"]]$data)
R> mn <- paste("mark", 1:ncol(simBC$geno[["ALL"]]$data), sep = "")
R> names(simBC$geno[["ALL"]]$map) <- mn
R> dimnames(simBC$geno[["ALL"]]$data)[[2]] <- mn
R> simBCs <- subset(simBC, chr = "ALL")
R> totmar(simBCs)
\end{Sinput}
\begin{Soutput}
[1] 54905
\end{Soutput}
\end{Schunk}
In \pkg{ASMap} the temporary omission of co-locating markers from an
unconstructed marker set can be achieved extremely efficiently using
\begin{Schunk}
\begin{Sinput}
R> simBCc <- pullCross(simBCs, type = "co.located")
R> totmar(simBCc)
\end{Sinput}
\begin{Soutput}
[1] 3501
\end{Soutput}
\end{Schunk}
When \code{type = "co.located"} the \code{pullCross()} makes use of the function
\code{findDupMarkers()} in \pkg{qtl} which has been explicitly written to ascertain
marker duplication. The \code{"geno"} component of the returned object
\code{simBCc} now contains less than 7\% of the original markers and
can be easily processed by \code{mstmap.cross()}. The co-located
markers, and their links to the markers remaining in the
unconstructed set, are retained to ensure that they can be pushed back into the
linkage map once the construction is complete (see Section
\ref{sec:pp}). This simple initial
diagnostic procedure now gives users the ability to construct very
large linkage maps without potential computational issues.

\section{Summary}

The \proglang{R} package \pkg{ASMap} provides an efficient suite of
tools for linkage map construction and diagnosis. The construction
functions utilise the source code of the MSTmap algorithm derived in
\cite{mst08} and can be used in various flexible ways to construct
linkage maps for large genetic marker data sets. The functions are also
extremely efficient with negligble loss of computational efficiency
compared to the MSTmap source code equivalent.

The package is under continuous development and updates will appear
through CRAN. In the short term, it is expected most of these developments will pertain
to increases in the efficiency of the linkage map construction work flow
through additional visual and numerical diagnostic functions. For
example, this includes a function that assists in the assignment and
alignment of linkage groups through the use of previously constructed
linkage maps. In the longer term, package updates may also include
functions that exploit future marker clustering and marker ordering
technology as they become available.

\section*{Acknowledgements}

The authors would like to sincerely thank Yonghui Wu, Stefano Lonardi
and Timothy Close for allowing us to manipulate and wrap the
MSTmap source code for use in the \pkg{ASMap} package. The authors
would also like to acknowledge and thank the anonymous referees whose
insightful comments allowed us to improve the manuscript.

\bibliography{jss2575}

\end{document}